\documentclass[12pt,tightenlines,eqsecnum,floats,aps,amsmath,amssymb,nofootinbib,prd]{revtex4}

\usepackage{amsmath,amscd}

\def\T{\mathcal{T}}
\def\w{\omega}

\def\H{\mathcal{H}}
\def\zb{\bar{z}}
\def\Hgq{\mathcal{H}}
\def\rgq{\rho^{\rm GQ}}
\def\H0{\mathcal{H}^{0}}
\def\Hgns{\mathcal{H}^{\rm GNS}}
\def\rgns{\rho^{\rm GNS}}

\def\rpol{\rho^{\rm pol}}

\def\R{\mathbb{R}}
\def\C{\mathbb{C}}

\def\i{{\rm i}}
\def\V{\mathcal{V}}
\def\Hpol{\mathcal{H}^{\rm pol}}
\def\Fpol{F^{\rm pol}}
\def\Vc{V^\C}
\def\th{\theta}
\def\partialx{\frac{\partial}{\partial x}}
\def\partialy{\frac{\partial}{\partial y}}
\def\Db{\overline{D}}
\def\Zb{\bar{Z}}
\def\WW{\mathcal{W}}
\def\W{{\rm W}}
\def\xt{x_\theta}
\def\yt{y_\theta}
\def\F{\mathcal{F}}
\def\v{{\rm v}}

\def\H{\mathcal{H}}
\def\Hpr{\mathcal{H}^{\rm pre}}
\def\A{\mathbf{A}}
\def\Tpol{\mathcal{T}^{\rm pol}}
\def\nabh{\nabla^{\mathcal{H}}}

\def\Tpol{\mathcal{T}^{\rm pol}}

\def\ub{\bar{u}}
\def\partialzu{\frac{\partial}{\partial z_u}}
\def\tr{{\rm tr}\,}
\def\trpol{{\rm tr^{pol}}\,}
\def\sb{{\rm SB}}
\def\pol{{\rm pol}}

\def\Hsch{\mathcal{H}^{\rm Sch}}
\def\emvs{e^{-{||\v||^2_u\over 2}}}
\def\Hfock{\mathcal{H}^{\rm Fock}}

\def\be{\begin{equation}}
\def\ee{\end{equation}}
\def\ba{\begin{eqnarray}}
\def\ea{\end{eqnarray}}

\begin{document}
\title{Polymer representations and geometric quantization}

\author{Miguel Campiglia} \email{miguel@gravity.psu.edu}
\affiliation{Institute for Gravitation and the Cosmos \& Physics
 Department,
 Penn State, University Park, PA 16802-6300, U.S.A.}

\begin{abstract} 
Polymer representations of the Weyl algebra of linear systems provide the simplest analogues of the representation used in loop quantum gravity. The construction of these representations is algebraic, based on the Gelfand-Naimark-Segal construction. Is it possible to understand these representations from a Geometric Quantization point of view? We address this question for the case of a two dimensional phase space.
\end{abstract}

\maketitle

\section{Introduction}
Consider a linear phase space $V$. Its simplest and best understood quantization is given by the Fock representation.  The quantization involves a choice of creation and annihilation operators. Let us denote by $u$ a parameter specifying such a choice, and let $\Hfock_u$ be the corresponding Fock space. Geometric quantization (GQ) \cite{woodhouse}  provides a way to obtain this space starting from the classical system: It is the space of square integrable complex polarized functions on $V$; $u$ enters as specifying the complex polarization\footnote{An alternative way  to understand $\Hfock_u$ from the classical space is as follows \cite{wald}:  $u$ is used to select a complex subspace $\H^1_u \subset V^\C$ on which the Hermitian form $-\i \w(\cdot,\bar{\cdot})$ is positive definite ($\w$ is the symplectic form on $V$ and the bar denotes complex conjugation). The Fock space is then given by $\Hfock_u=\sum_{n=0}^{\infty}(\H^1_u)^{\otimes n}$. Although this approach is restricted to linear systems, it has the advantage of being better suited for field theories. An extension of the present ideas to that context might need this viewpoint.}. 

Fock quantization however is not the final story; for instance, a different type of representation would be needed if one were to have a non-perturbative description of an interacting quantum field theory \cite{haag}. Non-Fock representations can also arise for kinematical reasons, as in the case of Loop quantum gravity \cite{biqg}, where the additional requirement of diffeomorphism covariance singles out a  representation \cite{lost}.  In both cases one takes an algebraic approach to the problem of quantization, where the focus is shifted to a chosen algebra associated to the classical system.

Let us illustrate the approach for the case of the Weyl algebra $\WW$ of our linear phase space $V$. The algebra is constructed from abstract generators $\W(f)$ labeled by linear functions $f:V \to \R$, with product rule 
 \be \label{weyl}
\W(f)\W(g)=e^{{\i \over 2}\{f,g\}}\W(f+g)
\ee
where $\{\cdot,\cdot \}$ is the Poisson bracket. States are then defined by abstract expectation values, $\langle \cdot \rangle: \WW \to \R$, known as positive linear functionals (PLF). The Hilbert space description is finally recovered by means of the so-called Gelfand-Naimark-Segal (GNS) construction, whose sole ingredient is one such PLF. 

The Fock quantizations provide examples of such PLFs,  given by the vacuum expectation values:
\be \label{uev}
\langle \W(f) \rangle_u = e^{-{||f||^2_u\over 2}}
\ee
where $||f||^2_u$ is obtained from the inner product induced by $u$. In this approach, the PLF (\ref{uev}) is used to \emph{define} the Fock representation. The space $\Hfock_u$ is then obtained from the GNS construction. Other choices of PLFs will give rise to representations that may or may not be unitarily equivalent to a Fock one. 

One may think the last possibility would only arise in field theory, since in finite dimensions the Stone-von Neumann theorem assures unitary equivalence of the representations of $\WW$. The theorem however only applies for representations on which the operators $\widehat{\W(f)}$ are continuous in $f$. If one gives up the continuity requirement,  non-Fock representations can be obtained \emph{even for finite dimensional systems} by considering PLFs such that $f \to \langle \W(f) \rangle$ is discontinuous; these are the so called polymer representations \cite{afw}. The reason to consider such PLFs is that they represent the Weyl algebra analogue of the Loop quantum gravity representation, where a similar discontinuity appears. In particular, they play a predominant role in Loop quantum cosmology  \cite{lqc}. The objective of the present work is to understand this non-standard quantization of finite-dimensional linear systems from the perspective of Geometric quantization. We hope this will provide a first step towards a study of the field theory situation \cite{cl}, where the relation between Fock and loop representations is a well studied subject \cite{lft}.

The paper is organized as follows.  In the next section we give an overview without going into details. We then proceed more systematically.  In section \ref{polgq} we introduce the space of polarizations and review basic notions of geometric quantization; in section \ref{gns} we introduce the polymer representations in a way geared towards our objective; finally in \ref{polypt} we bring together the topics of the previous two sections to understand how the polymer representations fit into the geometric quantization framework. We conclude in section \ref{conclusions}.

\section{Overview}\label{ov}
Let $V$ be two-dimensional, with linear coordinates $x$ and $y$ and symplectic structure $\w= dx \wedge dy$.
Consider the PLF \cite{afw}:
\be \label{polev}
\langle \W(a x+ b y) \rangle_\pol = \left\{ \begin{array}{ll} 1
& {\rm if} \quad x=0 \\
0 & {\rm otherwise}
\end{array}\right. .
\ee

Its GNS construction produces a Hilbert space $\Hpol$  whose elements are of the form
\be \label{polvec}
|\phi \rangle = \sum_x \phi(x) |x \rangle , \quad {\rm with } \quad ||\phi||_\pol^2= \sum_x |\phi(x)|^2 < \infty, 
\ee
where the $|x \rangle$ are normalized eigenvectors of $\hat{x}$ and $\phi(x)$ is a `function' with support on countably many points. Intuitively, it corresponds to a Schrodinger representation with a non-standard inner product in which the Dirac deltas became normalized.

In \cite{cvz}, the observation was made that the polymer PLF (\ref{polev}) can be obtained as a limit of Fock PLFs (\ref{uev}). Let us for simplicity look at the expectation values of the element $\W(x)$ and consider the following family of PLFs:
\be \label{cev} 
\langle \W(x) \rangle_d = e^{-\frac{x^2}{4 d^2}}.
\ee  
This is a particular instance of (\ref{uev}), with $d \in (0,\infty)$ playing the role of $u$. The observation is that, 
\be \label{plflimit}
\lim_{d \to 0} \langle \W(x) \rangle_d =\delta_{x 0} = \langle \W(x) \rangle_\pol,
\ee
so one recovers the polymer PLF (\ref{polev}). This way of obtaining the polymer space gives a first hint on to how make contact with GQ.  There, the Fock space associated to (\ref{cev}) is given by complex polarized functions with $d$ specifying the  polarization. From this viewpoint, $d = 0$ corresponds to a real polarization. This suggests we consider (\ref{polvec}) as providing a non-standard inner product for real polarized functions. Is this consistent with the GQ framework? Before addressing this question, let us make a small detour to discuss the polarized spaces of geometric quantization. 

There actually is a subtlety with the real polarization $d=0$: the corresponding functions are not normalizable. This issue is solved by the so-called metaplectic correction \cite{woodhouse}, in which the Hilbert spaces are constructed from half-forms rather than functions. This yields a well-defined inner product for the real polarized case (corresponding to the Schrodinger representation). It is in this framework that $d=0$ and $d>0$ stand on equal footing. 

The GQ framework also provides a way to compare the different $d$-spaces by means of a connection in the space of polarizations \cite{adpw}, that allows one to unitarily relate the different Hilbert spaces via parallel transport. For $d>0$, the corresponding map is nothing but a unitary Bogoliubov transformation between the Fock spaces. A non-trivial fact of this parallel transport is that it can be extended to the $d=0$ polarization \cite{kw}, \emph{provided} one includes the metaplectic correction (it is afterall the framework that allows one to deal with the $d=0$ space). In this case the map reproduces the Segal-Bargmann transform between the corresponding Fock and Schrodinger spaces. 

We would like to follow a similar strategy to relate the polymer space at $d=0$ with the Fock spaces at $d>0$. We  pointed out that,  in the metaplectic version of GQ, the Fock and Schrodinger spaces are related by the parallel transport \cite{kw}. We will see how,  \emph{without} the metaplectic correction, the Fock spaces are related to the \emph{polymer spaces} instead. This `exclusion' of metaplectic correction in the analysis is natural from the GNS perspective, since its role is irrelevant there: The PLFs (\ref{cev}) remain unaltered under the inclusion of the metaplectic term\footnote{The reason being that the Weyl algebra involves (exponentials of) operators associated to \emph{linear} functions, whose action is unaltered by the metaplectic correction.}. 

Let us now be more precise. The first thing to notice is that our situation is actually quite different from the one in \cite{kw}:  The space we have at $d=0$, $\Hpol$,  is unitarily inequivalant to the $d>0$ Fock spaces! In what sense can we compare them? Let us for the sake of simplicity identify (via Segal-Bargmann transform) the  $d>0$ spaces with  $\Hsch=L^2(\R,d x)$.  It becomes clear that there is no natural way to associate vectors in $\Hpol$ with vectors in $\Hsch$. For instance, the natural element in the Schrodinger space associated to  $|x\rangle \in \Hpol$ is given by the non-normalizable bra $(x|$  which lies outside $\Hsch$ \cite{afw}. On the other hand, if one looks at \emph{operators} on the corresponding spaces, one can do better. Let 
$|\phi \rangle \in \Hsch$ be given by a square integrable function $\phi(x)$, and  $|x\rangle \in \Hpol$ as before. Consider then the following operators\footnote{It is easy to check that both (\ref{polyop}) and (\ref{schop}) map normalized vectors to normalized vectors in their corresponding spaces; see also \cite{afw}.} 
\ba 
| \phi \rangle \otimes ( x|: &  \Hsch_x \to \Hsch_x \label{schop}\\
| x \rangle \otimes (\phi|: & \Hpol_x \to \Hpol_x  \label{polyop} ,
\ea   
where $( \phi|$ denotes contraction with the non-normalizable vector (in the polymer sense)  given by $\phi(x)$ \cite{afw}. These operators, defined on the polymer and Schrodinger spaces, are naturally in one to one correspondence. As we will see in section \ref{polypt}, the parallel transport implements this mapping between them.

Thus, we will actually need to work in a slightly different version of the framework in order to deal with operators rather than vectors. This resembles working with density matrices in quantum mechanics. In particular, the notion of a  unitary map is replaced by the notion of a trace-preserving map. In this language for instance, the unitary equivalence of the $d>0$ spaces follows from the trace-preserving property of the parallel transport. Conversely, the non-unitary equivalence of the Polymer space will result in a generic non trace-preservation of the parallel transport. We will however find a certain class of operators, namely those corresponding to (\ref{schop}), for which the parallel transport acts in a trace-preserving way, yielding the corresponding operators (\ref{polyop}). It is in this sense that we find a consistency of the polymer spaces with the geometric quantization framework.

\section{Space of polarizations and geometric quantization} \label{polgq}
The setting is as in the previous section, where  $V$ is a two-dimensional symplectic vector space with linear coordinates $x$ and $y$ and symplectic structure $\w=dx \wedge  dy$. We will use the conventions 
\be 
X_f:=-\partial_y f \ \partial_x + \partial_x f \ \partial_y
\ee
for the Hamiltonian vector of a phase space function $f$ (equivalently $df=-\w(X_f,\cdot)$),  and 
\be \label{pb}
\{f,g\}:=\w(X_f,X_g)=\partial_x f \ \partial_y g-\partial_y f \ \partial_x g
\ee
for the Poisson bracket.

The prequantum Hilbert space $\Hpr$ consist of square integrable functions $\psi: V\to \C$ with respect to the measure $\w/(2\pi)$. These functions are tied to the symplectic form by a covariant derivative with curvature $-\i \w$:
\be
\nabla_X \psi :=X(\psi)-\i\tau(X)\psi
\ee
where $\tau$ is such that $\w = d \tau$. A change in the symplectic potential $\tau \to \tau + d f$ is then compensated by a gauge transformation $\psi \to e^{\i f}\psi$. For concreteness we will work in the gauge given by
\be \label{tau}
 \tau:=(x dy-y dx)/2.
\ee
The Poisson algebra of real phase space functions is then represented by Hermitian operators 
\be\label{rho}
\rho(f)\psi:=-\i\nabla_{X_f}\psi+f\psi,
\ee
so that $[\rho(f),\rho(g)]=-\i \rho(\{f,g\})$.
To define a quantization one needs to introduce a polarization, which can be either real or positive complex. In our two dimensional case, the polarizations are given by one-dimensional complex subspaces of $\Vc$ on which the sesquilinear form $-\i \, \w(\cdot,\bar{\cdot})$ either vanishes (real case) or is positive-definite (complex case). Such spaces can be parametrized by the closed unit disk
\be
\Db :=\{ u\in \C ; |u| \leq 1 \}
\ee
when their generators are written in terms of a reference vector $Z_0=\tfrac{1}{\sqrt{2}}(\partialx-\i \partialy)$ as 
\be
Z_u :=  Z_0 + u \bar{Z}_0 .
\ee
One can then check that $-\i \w(Z_u,\bar{Z}_u)=1-|u|^2$ which is nonnegative for $u \in \Db$. To see how the disk boundary corresponds to real polarizations, we take $u=e^{\i \th}$ and rescale the vector as   
\ba \label{zth}
Z_\th & := & \frac{1}{\sqrt{2}} e^{-\i \th/2}\, Z_{e^{\i \th}} \\
& = & \cos \th/2 \partialx -\sin \th/2 \partialy ,
\ea
which provides a parametrization of the one-dimensional real vector spaces of $V$.

When $|u|<1$ the complex polarization induces a complex structure on $V$. The corresponding holomorphic coordinates can be taken to be
\be \label{zu}
z_u:=\frac{z_0-\ub \ \zb_0}{\sqrt{1-|u|^2}}
\ee
where $z_0 \equiv \tfrac{1}{\sqrt{2}}(x+ \i y)$. The normalization factor makes the derivative
\be
\partialzu= \frac{Z_u}{\sqrt{1-|u|^2}}  
\ee
normalized so that $\w = \i d z_u \wedge d \zb_u$. In the quantum theory this will correspond to canonically normalized Fock operators. Finally, the symplectic potential (\ref{tau}) takes the form
\be 
\tau=\i (z_u d \zb_u - \zb_u d z_u)/2.
\ee

Given $u \in \Db$, let us denote by $\V_u$ the space of polarized functions satisfying the condition
\be \label{polsec}
\nabla_{\Zb_u}\psi=0.
\ee

When $|u|<1$, solutions to (\ref{polsec}) are of the form
\be \label{cpolsec}
\psi(x,y)=\phi(z_u)e^{-{1\over 2}|z_u|^2},
\ee 
and by restricting to square integrable functions one obtains the Hilbert space $\H_u \subset \Hpr$ of normalizable complex polarized functions. There exists a natural orthogonal basis given by $\phi=\{1,z_u,z_u^2,\ldots\}$ and one recovers the Fock space corresponding to creation and annihilation operators associated to (\ref{zu}).

When $u=e^{i \theta}$, the solutions of (\ref{polsec}) are given by
\be \label{rpolsec}
\psi(x,y)=e^{\i {\xt \yt \over 2}}\phi(\xt),
\ee 
where 
\ba
\xt & := & x \sin\tfrac{\th}{2} + y \cos \tfrac{\th}{2}  \\ \label{xt}
\yt & := &-x \cos \tfrac{\th}{2}  + y \sin \tfrac{\th}{2} 
\ea
are canonically conjugated variables. Apart from the phase factor in (\ref{rpolsec}) (which comes from the gauge choice (\ref{tau})), the space is given by  functions $\phi(\xt)$ on which $\rho(\xt)$ and  $\rho(\yt)$ act in the standard way:
\ba
\rho(\xt) e^{\i {\xt \yt \over 2}}\phi(\xt) & = & e^{\i {\xt \yt \over 2}} \xt \phi(\xt) \label{rhoxt} \\ 
\rho(\yt) e^{\i {\xt \yt \over 2}}\phi(\xt) & = & e^{\i {\xt \yt \over 2}} \i \frac{\partial}{\partial \xt} \phi(\xt) \label{rhoyt}. 
\ea
One seems to be recovering the Schrodinger representation in the $\xt$ variable, except for the caveat that the GQ inner product is given by integration over $V$ with measure $d x d y = d x_\th d y_\th$.  At this point one could fix this by introducing a new prescription for the inner product on real polarized functions, in such a way that the operators $\phi(\xt)$ and $\rho(\yt)$ are Hermitian. For instance, one could declare it to be given by integration over $d x_\th$, thus recovering the Schrodinger representation. This is however not satisfactory if one wants a unified framework for both real and complex polarized spaces. The issue is nicely solved by the metaplectic correction\footnote{Which at the same time fixes the quantization of the quadratic operators, and the corresponding representation of the symplectic group \cite{woodhouse}.}. We will see that the polymer representations provide an alternative prescription in the absence of the metaplectic correction.

Let us finish this section by mentioning that we will only be interested in representations of real linear phase space functions $f \in V^*$, in which case the operators (\ref{rho}) leave the polarized spaces invariant and so are automatically well defined in $\V_u$.

\section{Polymer representations as limits of GNS representations}\label{gns}
In the algebraic approach to quantization, the focus is shifted to a Poisson subalgebra of phase space functions. In linear systems it is natural to consider the Heisenberg algebra of linear functions, or alternatively its exponentiated version, the Weyl algebra $\WW$.\footnote{In studying their representations the latter is mathematically simpler since it involves bounded operators.} In our system,  $\WW$ is the *-algebra consisting of linear combinations of abstract elements $\W(\v)$ with $\v\in V^*$, subject to the product rule
\be
\W(\v_1)\W(\v_2)=e^{{\i \over 2}\{\v_1,\v_2\}}\W(\v_1+\v_2)
\ee 
($\{\v_1,\v_2\}\equiv \w^{-1}(\v_1,\v_2)$ is the Poisson bracket (\ref{pb})) and star operation  $\W^*(\v):=\W(-\v)$. Once the algebra is selected, states are abstractly defined as PLFs: Linear functionals $\F:\WW \to \R$ obeying the positivity condition $\F(W^* W) \geq 0$ for all possible linear combinations of the generators, $W=\sum_n a_n \W(\v_n) \in \WW$. The passage to the the standard description in terms of operators on a Hilbert space is given by the GNS construction  (see for instance section 4.5 of \cite{wald}). 

The Hilbert spaces $\Hgq_u$ ($|u|<1$) introduced in the previous section carry a unitary representation of the Weyl algebra given by
\be
\rgq_u(\W(\v)) \equiv e^{\i \rho(\v)} : \Hgq_u \to \Hgq_u,
\ee
($\rho(\v)$ is the GQ operator (\ref{rho})), and so any normalizable vector in $\Hgq_u$ defines a PLF on $\WW$ (given by its expectation values). Let us focus on the `simplest' vector given by $\phi=1$ in (\ref{cpolsec}), and let $\F_u$ be its corresponding PLF. From the linearity property such a functional is fully specified by its value on the algebra generators $\F_u(\W(\v))=:F_u(\v)$ and a simple calculations leads
\ba 
F_u(\v) & = & \int  e^{-{1\over 2}|z_u|^2}e^{\i \rho(\v)}e^{-{1\over 2}|z_u|^2} \frac{dx dy}{2 \pi}\\
& = & \emvs \label{Fu},
\ea 
where $||\v||^2_u$ is given by the inner product on $V^*$ induced by the complex structure (see appendix \ref{gnslimit}). Let us now `forget' where (\ref{Fu}) came from and take $F_u(\v)=\emvs$  as defining an abstract PLF on $\WW$. The corresponding GNS construction gives rise to a unitary irreducible representation (UIR) of the Weyl algebra $(\rgns_u,\Hgns_u)$ which is in one to one correspondence with  ($\rgq_u$,$\H_u$).  But now we can make sense of the function (\ref{Fu}) for points sitting at the disk boundary. Writing $u= r e^{\i \th}$ and taking the $r \to 1$ limit, one obtains (see appendix \ref{gnslimit}),
\be \label{fpol}
\Fpol_\th(\v):=\lim_{r \to 1} \emvs= \left\{ \begin{array}{ll} 1
& {\rm if} \quad \v(Z_\th)=0 \\
0 & {\rm otherwise},
\end{array}\right.
\ee
where $Z_\th$ is given in (\ref{zth}) and $\v(Z_\th)=0 \iff \v(x,y)\propto x_\th$. One can check that the corresponding functional $\F^\pol_\th:\WW \to \R$ satisfies the positivity condition thus constituting a PLF.  The GNS construction produces then an UIR $(\rpol_\th,\Hpol_\th)$ of the Weyl algebra; these are the so called polymer representations \cite{afw}.\footnote{There is another polymer representation that has been considered in the literature which does not fall into the family (\ref{fpol}): It corresponds to a function $F(\v)$ which vanishes everywhere except at $\v=0$ \cite{hp}. We have not studied this representation in detail, but it seems that a connection with GQ would be at the prequantum level, since such PLF carries no information of polarization.}

Now, in our present finite-dimensional setting, representations of the Weyl algebra are restricted by the Stone-von Neumann theorem, which tell us that all UIRs $\rho$ for which the map $\v \to \rho(\W(\v))$ is continuous are unitarily equivalent. This is the situation for the representations $\rgq_u$, and so, up to unitary equivalence, the whole open disk corresponds to a single representation (in section \ref{polypt} we will see this result from the GQ perspective). On the other hand, all the polymer representations $\rpol_\th$ are inequivalent. The theorem does not apply to them because the operators $\rpol_\th(\W(\v))$ are not continuous\footnote{In particular one cannot recover a representation of the Heisenberg algebra by differentiating the Weyl operators. From this perspective they would be regarded as unsatisfactory; we recall however that the interest in the polymer representations lies in their role as analogues of the loop representation, for which the corresponding Heisenberg algebra is also not represented.}. 

Let us now summarize the main properties of the polymer representations. For each $\th$, condition $\v(Z_\th)=0$ determines a preferred direction along which the operators $\rpol_\th(\W(\v))$ are continuous, i.e.,   $t \to \rpol_\th(\W(t  \xt))$ is continuous in $t$.  Furthermore, one can differentiate along this direction to obtain an operator $\rpol(\xt)$  associated to $\xt$. Its eigenvectors provide a normalizable basis, with respect to which vectors are given by `functions' $\phi(\xt)$ with support on a countable number of points. In such a representation the inner product takes the form,
\be \label{polynorm}
||\phi||^2=\sum_{x_\th \in \R}|\phi(x_\th)|^2,
\ee
and the action Weyl algebra is given by  
\ba
\rpol_\th(\W(t  \xt)) \phi(\xt) & = & e^{i t \xt}\phi(\xt) ,\label{rhoext}\\
\rpol_\th(\W(t  \yt)) \phi(\xt) & = & \phi(\xt-t) \label{rhoeyt}.
\ea
Finally, there exists a dual representation in which vectors are represented by quasiperiodic functions of $\yt$. In terms of the previous basis it is given by $\tilde{\phi}(\yt):=\sum_{\xt}\phi(\xt) e^{-\i \xt \yt}$, and the inner product takes the form 
\be \label{polynorm2}
||\phi||^2= \int \overline{d \yt} |\tilde{\phi}(\yt)|^2  := \lim_{L \to \infty} (2 L)^{-1}\int_{-L}^L d \yt |\tilde{\phi}(\yt)|^2 . 
\ee

Going back to GQ, we see that (\ref{rhoext}),(\ref{rhoeyt}) correspond to the exponentiated versions of (\ref{rhoxt}),(\ref{rhoyt}), and so we could think of the polymer spaces as providing an inner product for the real polarized functions\footnote{The inner product we obtain is not the one expected from the analogy with loop quantum gravity. If $x$ where to represent the connection variable, we would expect a prescription $\int \overline{d x}|\phi(x)|^2$ (\ref{polynorm2}) instead of  $\sum_{x \in \R}|\phi(x)|^2$ (\ref{polynorm}) for the $x$-polarized functions.}. The prescription corresponds to the formal substitution $\int d \xt d \yt \to \sum_{\xt} \int \overline{d \yt}$ for the integration of such functions, and so it amounts to a change of the original prequantum  measure.

\subsection{Alternative viewpoint} \label{altvp}

Let us finish by giving a more intuitive version of the construction.  For each polarization associated to a point in the open disk $u \in D$ ,  we have a `preferred' polarized function given by $e^{-{1\over 2}|z_u|^2} \in \H_u$ (the `vacuum state'). It is normalized to unity, with respect to the $dx dy/(2 \pi)$ measure on $V$. We now use this family of $u$-polarized functions to \emph{define} an inner product on the real polarized spaces as follows. We first extend the functions to the boundary points by setting $u=r e^{\i \th}$ and taking the $r \to 1$ limit. In a similar way as (\ref{fpol}), one can  show that
\be \label{deltaxt}
\lim_{r \to 1} e^{-{1\over 2}|z_{u}|^2}=\delta_{\xt 0},
\ee  
which provides a `vacuum' for the $u=e^{\i \th}$ polarized space\footnote{Since $\delta_{\xt 0}=e^{\i {\xt \yt \over 2}} \delta_{\xt 0}$, this `function' is of the form (\ref{rpolsec}) (one would actually need to re-express condition (\ref{polsec}) in an exponentiated version, in order to make sense for the type of `functions' featuring in the polymer space; the solutions however will still be given by  (\ref{rpolsec}).}. Finally, by requiring (\ref{deltaxt}) to be normalized, (and that the Weyl operators act unitarily), one recovers the inner product (\ref{polynorm}).  We thus end up with a family of Hilbert spaces parametrized by the closed disk $\Db$, and on each Hilbert space there is a normalized `vacuum', which corresponds to the null eigenvector of the `annihilation' operator $\rho(\zb_0- u  z_0)$. 

Unfortunately, the normalization of the polymer inner product emerging from this picture differes from the one suggested by the parallel transport  (see section \ref{sqrt2}). We will thus not rely on this picture in the subsequent analysis. 

\section{Parallel transport and polymer representations}  \label{polypt}

In GQ, a way to understand the relation between the spaces $\Hgq_u$ is by viewing them as fibers of a Hilbert space bundle $\H \to D$, where $D$ is the open unit disk. Since $\H$ is a subbundle of the bundle $\Hpr \times D$, the trivial connection in the latter induces a  connection  $\nabh:=d+A$ in the former given by \cite{adpw,kw,wu}:
\be \label{connection}
A=  -{d \bar{u}\over 2(1-|u|^2)}\nabla^2_{z_u},
\ee
where $\nabla_{z_u}\equiv \nabla_{\partial/\partial_{z_u}}$. By construction the connection is unitary (see also appendix \ref{parallelt}). Its relevance comes from the additional fact that its curvature is proportional to the identity. This allows one to have (up to a phase) a canonical unitary map between the $\H_u$ spaces given by the parallel transport, which in turn can be used to provide an explicit realization of the Stone-von Neumann theorem for the $(\rgq_u,\H_u)$ representations. Finally, it is worth mentioning that the inclusion of the metaplectic correction removes the phase ambiguity, i.e., the corresponding connection is exactly flat \cite{kw,woodhouse}.

So far the connection and parallel transport dealt with points in the disk interior $|u|<1$. In \cite{kw} Kirwin and Wu showed how it is possible to extend the parallel transport to points in the disk boundary by taking limits along geodesics in $D$.\footnote{The metric on $D$ is $ds^2= 4(1-|u|^2)^{-2}d u d\bar{u}$ \cite{wu}.}  In their construction the metaplectic correction is essential, showing again how this term allows one to treat real and complex polarized spaces on an equal footing.

Now, in the GNS-type construction of the previous section, the metaplectic correction played no role. Even if included, the expectation value (\ref{Fu}) remains unchanged. Our proposal then is to study the parallel transport on the non-metaplectic bundle, and attempt to relate it with the polymer representations when extended to the disk boundary. 

The strategy however has to be different from the one in \cite{kw}, since we are now associating inequivalent representations to the disk boundary: There cannot be any unitary map between the spaces $\Hpol_\th$ and $\H_u$. What should we attempt to relate then? It turns out that the best one can do is to relate certain operators defined on both spaces. These are the type of operators (\ref{polyop}) and (\ref{schop}) discussed earlier.

Following this idea, we shift attention from the bundle $\H$ to the bundle $\T$ of finite trace operators. Let us be more specific. Recall the space $\H_u$ is defined by square integrable functions in $\V_u$ (\ref{polsec}). Let us similarly define the space $\T_u$ as given by elements in $\V_u \otimes \bar{\V}_u$ (the bar denotes complex conjugation), i.e., functions $O:V \times V \to \C$ obeying\footnote{The superscripts denote dependence on the first and second copy of $V$ and $\bar{\nabla}_{X} \equiv X+i\tau(X)$} 
\be \label{polop}
\nabla_{\bar{Z}_u^1}O=0=\bar{\nabla}_{Z_u^2}O,
\ee
such that the operator  $(O \psi)(x^1,y^1) = \int O(x^1,y^1;x^2,y^2) \psi(x^2,y^2) {dx^2 dy^2 \over 2 \pi}$ has finite trace and maps square integrable  functions to square integrable functions. Condition (\ref{polop}) implies $O$ is given by a function $T(z_u^1,\bar{z}_u^2)$ as
\be \label{O}
O(x^1,y^1;x^2,y^2)  =  e^{-{1\over 2}(|z_u^1|^2+|z_u^1|^2)} T(z_u^1,\bar{z}_u^2) 
\ee
in terms of which the trace is given by
\be
\tr T := \tr O = \i \int T(z_u,\bar{z}_u) e^{-|z_u|^2} {d z_u d \zb_u \over 2 \pi}. 
\ee
  
Now, the connection (\ref{connection}) induces a connection  $\A$ on $\T$ as follows. First, $A$ makes sense as a connection on the bundle of polarized functions $\V \to D$, which by imposing Leibniz rule, can be extended to $\V \otimes \bar{\V} \supset \T$   yielding,
\be \label{connT}
\A = - {1\over 2(1-|u|^2)} (d \bar{u} \nabla^2_{z_u^1} + d u  \bar{\nabla}^2_{\zb_u^2} ). 
\ee
From the unitarity property of the original connection it follows that  $\A$ can be restricted to $\T$, and that its parallel transport is trace-preserving. Notice that the original bundle $\H$  can be mapped inside $\T$ by tensoring with itself: $\phi(z) \to  \phi(z^1) \bar{\phi}(\bar{z}^2)$, which corresponds to working in a density matrix framework\footnote{There, the notion of unitary map is translated into the notion of trace-preserving map. From that perspective, the bundle $\T$ represent an alternative to the bundle $\H$ which is better suited to make contact with the polymer spaces.}. Finally, one can verify that the connection (\ref{connT}) has vanishing curvature, reflecting the absence of global phase factors when going to a density matrix description $\phi(z) \to  \phi(z^1) \bar{\phi}(\bar{z}^2)$.

Let us summarize the situation so far. We are trying to relate the $\H_u$ complex polarized spaces in the disk interior with the polymer spaces $\Hpol_\th$ sitting at the disk boundary, within the GQ framework. An appropriate tool to do so is by means of the parallel transport, along the lines of \cite{kw}. In order to deal with the fact that the polymer spaces are unitarily inequivalent to the GQ ones, we propose working in the bundle $\T \to D$ of finite trace operators, as opposed to standard bundle $\H \to D$. The unitary inequivalence of both spaces will be reflected in a generic non-trace-preservation of the parallel transport. However we expect operators of the type (\ref{polyop}) and (\ref{schop}) to be mapped to each other. 

Before studying the parallel transport in $\T$, let us introduce the spaces at the disk boundary. Given $u= e^{\i \th}$ consider now the space $\Tpol_\th$ of finite trace operators on $\Hpol_\th$. They can be thought of as given by `functions' $O:V \times V \to \C$ of the form
\be \label{opol}
O   =  e^{\i {\xt^1 \yt^1 \over 2}-\i {\xt^2 \yt^2 \over 2}} T(\xt^1,\xt^2),
\ee
obeying,
\be \label{trpol}
\trpol T := \frac{1}{\sqrt{2}} \sum_{\xt}T(\xt,\xt)< \infty
\ee
and such that\footnote{We are defining the action of the operator in a transposed way in order to simplify the discussion later.}
\be \label{tpol}
(T \phi)(x^2)= \sum_{x^1} T(x^1,x^2)\phi(x^1)
\ee
has finite polymer norm for normalizable $\phi$. The $1/\sqrt{2}$ factor in the definition of the trace (\ref{tpol}) is  needed in order to have the trace-preservation result (in section (\ref{sqrt2}) we will comment on the fact that this normalization differes from the one in (\ref{polynorm})).  We now have all the elements to study the parallel transport and its $|u| \to 1$ limit. 

Following \cite{kw}, we will consider approaching the boundary along curves whose final direction is perpedincular to the boundary\footnote{This is equivalent to considering the limit along the geodesics of $D$.}. This is equivalent to consider the limit $r \to 1$ of $u=r e^{\i \th}$, which also agrees with the way the limit was taken in section \ref{gns}. To simplify notation, let us take $\th=\pi$ in which case $x_\th=x$ and $y_\th=y$ (we will later translate the result to arbitrary $\th$). As in \cite{kw} we take the parametrization $r = \tanh t$. The corresponding holomorphic coordinates are given by
\be \label{zt}
 z_t:= z_{(u=-\tanh t)}={1\over \sqrt{2}}(e^{t}x+\i e^{-t}y).
\ee
We will consider the parallel transport $P_t: \T_0 \to \T_t$ from the initial fiber at $t=u=0$ to the fiber $\T_t:=\T_{u=-\tanh t}$. Our goal is to study the $t \to \infty$  limit of $P_t$ and its relation with the space $\Tpol_x:=\Tpol_{\th=-\pi}$.

Before going on, let us recall some results from \cite{kw} regarding the parallel transport in $\H$. Consider an initial state $\psi_0=\phi(z_0)e^{-{|z_0|^2\over 2}} \in \H_0$ and 
let $\psi_t=\phi_t(z_t)e^{-{|z_t|^2\over 2}}$ be its parallel transport along $u=-\tanh t$ (see appendix \ref{parallelt}). For finite $t$, the parallel transport reproduces the Bogoliubov transformation between the $z_0$ and $z_t$ Fock representations. In order to obtain a  nontrivial $t \to \infty$ limit one needs to include the metaplectic correction, which amounts using half-forms instead of functions. It turns out that, along our particular curve, an initial half-form $\tilde{\psi}_0=\psi_0 \sqrt{d z_0}$ is parallel transported to $\tilde{\psi}_t=\psi_t \sqrt{d z_t}$ where $\psi_t$ is given by the parallel transport in $\H$. The limit \cite{kw}:
\be \label{limitkw}
\lim_{t \to \infty} \phi_t(z_t)e^{-{1\over 2}|z_t|^2}\sqrt{d z_t} = e^{\i {x y \over 2}} \phi^{\sb}(x) \sqrt{d x} 
\ee 
gives a well defined real polarized half-form and the resulting map reproduces the unitary Segal-Bargmann transform:\footnote{The normalization factor in (\ref{SB}) corresponds to a  measure  $dx/\sqrt{2\pi}$ for the real space.}
\be \label{SB}
\phi^{\sb}(x'):= 2^{1/4}\int \frac{dx dy}{2 \pi} \phi(z_0)e^{-|z_0|^2} e^{-{\bar{z}_0^2\over 2}+\sqrt{2}x'\bar{z}_0-{x'^2\over 2}}.
\ee

The way the limit (\ref{limitkw}) works is that, for large $t$, the asymptotic behavior of the individual terms are \cite{kw}
\ba
\sqrt{d z_t} & \sim & 2^{-1/4}e^{t/2}\sqrt{d x} \label{sqdzt}\\
\phi_t(z_t)e^{-{1\over 2}|z_t|^2} & \sim & 2^{1/4}e^{-t/2}e^{\i {x y \over 2}} \phi^{\sb}(x) \label{limsb}
\ea
(the first relation can easily be seen from (\ref{zt}). From (\ref{limsb}) we see that the parallel transport of the non-metaplectic functions vanish in the $t \to \infty$ limit.   The metaplectic correction provides a $e^{t/2}$ `missing' factor for the limit to be non-trivial.

Let us now look at an initial function corresponding to the $x'$ eigenstate of the $\rho(x)$ operator:
\be \label{phix0}
\psi^{x'}_0=\phi^{x'}(z_0)e^{-{1\over 2}|z_0|^2}:= e^{-{1\over 2}|z_0|^2}e^{-{z_0^2\over 2}-{x'^2\over 2}+\sqrt{2}x'z_0}.
\ee
Eventhough $\psi^{x'}_0 \in \V_0$ is not normalizable, its parallel transport in $\V$ is well defined. It is given by (see appendix \ref{parallelt}): 
\ba 
\psi^{x'}_t=\phi^{x'}_t(z_t) e^{-{1\over 2}|z_t|^2} &=& e^{-{|z_t|^2\over 2}}e^{t/2} e^{-{z_t^2\over 2}-{x'^2\over 2}e^{2 t}+\sqrt{2}x'z_t e^{t}} \label{phix}\\
& = &  e^{\i y (x'-x/2)} e^{t/2} e^{-{e^{2 t}\over 2}(x-x')^2} \label{delta}
\ea
and again satisfies  $\rho(x) \psi^{x'}_t= x'\psi^{x'}_t$.  Notice that, with an additional $e^{t/2}$ factor from the metaplectic correction, the $t \to \infty$ limit reproduces the Dirac delta corresponding to the Schrodinger representation of the $x'$ eigenstate. 

By looking at (\ref{limsb}) and (\ref{delta}), we now see which elements in $\T$  have a non-trivial $t \to \infty$ limit. Consider then the initial condition $O_0=e^{-{1\over 2}(|z_0^1|^2+|z_0^1|^2)}T_0(z_0^1,\zb_0^2)\in \T_0$ given by 
\be \label{t0}
 T_0(z_0^1,\zb_0^2) =\phi(z_0^1)\bar{\phi^{x'}}(\bar{z}_0^2),
\ee
with $\phi^{x'}$ as before and $\phi(z_0)e^{-{1\over 2}|z_0|^2} \in \H_0$ . The trace is 
\be \label{trt0}
\tr T_0 = 2^{-1/4} \phi^{\sb}(x')
\ee
as can be seen from (\ref{phix0}) and (\ref{SB}). The parallel transport is given by the product of the parallel transported functions: $T_t=\phi_t(z_0^1)\bar{\phi^{x'}}_t(\bar{z}_0^2)$. Using (\ref{limsb}) and (\ref{delta}) we conclude that the limit
\be
\lim_{t \to \infty} e^{-{1\over 2}(|z_t^1|^2+|z_t^1|^2)} T_t(z_t^1,\bar{z}_t^2)  =  e^{\i {x^1 y^1 \over 2}-\i {x^2 y^2 \over 2}} 2^{1/4}\phi^{\sb}(x^1)\delta_{x^2 x'} \in \Tpol_x \label{limittpol}
\ee 
gives a well defined element of  $\Tpol_x$ (defined in equations (\ref{opol}) to (\ref{tpol})). Furthermore, the trace of the resulting operator in $\Tpol_x$ gives
\be \label{trtpol}
\frac{1}{\sqrt{2}}\sum_{x} 2^{1/4} \phi^{\sb}(x)\delta_{x x'} = 2^{-1/4}\phi^{\sb}(x'),
\ee
in agreement with (\ref{trt0}). In this way, the parallel transport implements the map (\ref{schop}) $\to$ (\ref{polyop}).

The same trace preservation holds for linear combinations of operators of the form (\ref{t0}). Let us denote by $(\T_x)_0 \subset \T_0$ the space of such operators (i.e. elements of the form $T_0=\sum_{n} \phi^n_0 \otimes \bar{\phi}^{x_n}_0 $). The parallel transport maps this space to the corresponding subspace of $\Tpol_x$ given by elements of the form $\sum_{n}\phi(x_1)\delta_{x_2 x^n}$. This picture extends to any initial point $u \in D$\footnote{One could consider curves going from $u$ to $0$ and then using the previous construction; recall the connection $\A$ is flat and so results are insensitive to the choice of curve; the limit to the boundary however has to be along curves whose final direction is perpendicular to the boundary.} and the collection of all such subspaces can be represented by a bundle  $\T_x \subset \T$. 

Similarly, associated to any boundary point $u=e^{\i \th}$ there is a subbundle $\T_\th \subset \T$ for which the parallel transport can be extended to the boundary point $u=e^{\i \th}$ in a way that the trace is preserved. 

Let us finish with the observation that by working with slightly more general elements one could consider curves connecting two boundary points. Let us illustrate the situation with $u=-\tanh t$ but now considering both limits $t \to \pm \infty$, which can then be thought as relating the polymer spaces $\Tpol_x$ and $\Tpol_y$ associated to $u=\mp 1$. From the previous analysis one can guess which elements will have non trivial limit in both directions. They correspond to:
\be \label{txy0}
T^{x'y'}_0:=\phi^{y'}(z_0^1)\bar{\phi}^{x'}(\bar{z}_0^2)
\ee
where $\phi^{y'}$ is an eigenfunction of $\rho(y)$ analogous to (\ref{phix0}). 
 Notice that, since $\phi^{y'}$ is not normalizable, (\ref{txy0}) does not yield an operator in $\T_0$. It does however have a finite trace
\be \label{trxy}
\tr T^{x'y'}_0=2^{-1/2}e^{-\i x' y'}
\ee
and a well defined parallel transport 
\be
T^{x'y'}_t= \phi^{y'}_t \bar{\phi}^{x'}_t,
\ee
given by equation (\ref{phix}) and the analogous one for $\phi^{y'}$:  
\ba
\phi^{y'}_t & = & e^{-t/2}e^{-{|z_t|^2\over 2}}e^{{z_t^2\over 2}-{y'^2\over 2}e^{-2 t}-\i\sqrt{2}y'z_t e^{-t}} \\
& = & e^{\i x y /2} e^{-t/2} e^{-\i y' x} e^{-{e^{-2 t}\over 2}(y-y')^2} \label{deltay}.
\ea 
By bringing together (\ref{delta}) and (\ref{deltay}), we find the following limits of the parallel transport 
\be
 e^{-{1\over 2}(|z_t^1|^2+|z_t^1|^2)} T^{x'y'}_t(z_t^1,\bar{z}_t^2) \to \left\{\begin{array}{lll}
e^{\i {x^1 y^1 \over 2}-\i {x^2 y^2 \over 2}} e^{-\i y' x^1} \delta_{x^2 x'} &\in \Tpol_x \ {\rm when} & t \to \infty \\
e^{-\i {x^1 y^1 \over 2}+\i {x^2 y^2 \over 2}} e^{-\i x' y^2} \delta_{y^1 y'} &\in \Tpol_y \ {\rm when} & t \to -\infty \end{array}\right.
\ee
Furthermore, the corresponding traces in the polymer spaces agree with (\ref{trxy}).

\section{Conclusions and Outlook} \label{conclusions}
The motivation for studying the polymer representations is that they provide the simplest analogues of the representation featuring in loop quantum gravity.  A natural question to ask is how these representations are related with more established quantization schemes, as for instance the one provided by geometric quantization. 

Here we focused on the simplest possible setting of a two dimensional phase space.  In section \ref{gns} we  argued that, from a geometric quantization perspective, the polymer representations should be associated to real polarized spaces (with a non-standard inner product). We then sought for consistency on this viewpoint by attempting to relate, via parallel transport, the polymer spaces with the complex polarized spaces of geometric quantization (section \ref{polypt}). Since the spaces being related are unitarily inequivalent, all we could say was that the parallel transport is trace-preserving  when restricted to certain type of operators, namely those of the type (\ref{schop}) and (\ref{polyop}) (we did not study what the situation is for arbitrary operators, beyond the fact that those given by tensoring two normalizable vectors in $\H_u$ have a vanishing parallel transport limit). These results hold for the parallel transport corresponding to the  non-metaplectic connection. 

An aspect of interest left for future work is the inclusion of constraints and corresponding gauge symmetries. It may be possible a situation in which the gauge invariant Hilbert spaces are unitarily equivalent, even though the underlying kinematical spaces are not. The simplest version of gauge symmetry that can be considered for our two dimensional phase space consists of discrete translations (see section 1.c of \cite{adpw}). This example is well understood \cite{theta}, and it would be interesting to see if those results could be obtained as a reduction of underlying Fock and polymer kinematical spaces.  

Finally, we would like to move beyond finite dimensions and study field theories with diffeomorphisms included in the gauge group. This is after all the context where loop-type representations arise.

\section*{Acknowledgments} I am indebted to Alok Laddha, for suggesting the topic and providing much guidance throughout this work. I would like to thank Abhay Ashtekar, Adam Henderson, Casey Tomlin and Artur Tsobanjan for discussions and comments. This work
was supported in part by the NSF grant PHY0854743 and the Eberly research funds of Penn State.

\appendix 
\section{Limit (\ref{fpol})}\label{gnslimit}
To understand the limit $r \to 1$ of $\emvs$, with $u=r e^{\i \th}$, let us write $\v$ as
\be \label{vab}
 \v=a\ d \xt + b\ d \yt.
\ee
One can then check that the norm on $\v$ induced by the complex structure is given by
\be
||\v||^2_u= a^2 \frac{1-r}{1+r}+b^2\frac{1+r}{1-r}.
\ee
Thus, in the $r \to 1$ limit the norm diverges unless $b=0$, in which case the limit is zero. This translates into 
\be \label{limitapp}
\lim_{r \to 1} \emvs= \left\{ \begin{array}{ll} 1
& {\rm if} \quad b=0 \\
0 & {\rm otherwise}.
\end{array}\right.
\ee
Finally, from (\ref{zth}) and (\ref{vab}) one can check that $b=-\v(Z_\th)$, thus obtaining (\ref{fpol}). 

\section{Parallel transport} \label{parallelt}
The parallel transport equation in $\H$ along a curve $u(t)\in D$ is given by 
\be \label{pt1}
\partial_t \psi -\frac{\dot{\ub}}{2(1-|u|^2)}\nabla^2_{z_u}\psi =0.
\ee
Writing $\psi=\phi(z_u,t)e^{-{1\over 2}|z_u|^2}$, equation (\ref{pt1}) is equivalent to
\be \label{pt2}
\partial_t \phi +\frac{1}{2(1-|u|^2)}\left( \dot{u} z_u^2-\dot{\ub}\partial^2_{z_u}+2(\ub \dot{u}-u \dot{\ub})z_u \partial_{z_u} \right) \phi=0.
\ee
Equation (\ref{pt2}) can be thought of as a Schrodinger equation with a time-dependent Hamiltonian. Recall that in the Fock language $z_u$ and  $\partial_{z_u}$ are creation and annihilation operators respectively, from which it becomes clear that the `Hamiltonian' in \ref{pt2} is Hermitian, and thus the resulting `evolution' unitary. 

Specializing to the case $u=-\tanh t$, equation (\ref{pt2}) becomes
\be \label{pt3}
\partial_t \phi -\tfrac{1}{2} z_t^2 \phi +\tfrac{1}{2}\partial^2_{z_t}\phi=0.
\ee
It is then easy to verify that
\be
\phi^{x'}_t(z_t) = e^{-{z_t^2\over 2}-{x'^2\over 2}e^{2 t}+\sqrt{2}x'z_t e^t+t/2} 
\ee
satisfies (\ref{pt3}).

\section{Normalizations in (\ref{polynorm}) and (\ref{trpol})} \label{sqrt2}
The agreement of the traces (\ref{trt0}) and (\ref{trtpol}) was met by using a particular normalization in the definition of the polymer trace (\ref{trpol}). With this normalization however, the picture from the discussion in section (\ref{altvp}) gets altered. Namely, the    `vacua'  $\delta_{\xt 0} \in \Hpol_\th$ will no longer be normalized, but have a norm equal to $1/\sqrt{2}$.  It is thus not possible to satisfy both i) normalization of $\delta_{\xt 0}$ and ii) trace preservation.

\end{document}